\documentclass[12pt,preprint]{aastex} 
\usepackage{graphics} 
\input epsf
\begin{document} 

\title{Dusty Acoustic Turbulence in the Nuclear Disks
of two LINER Galaxies NGC 4450 and NGC 4736} 

\author{Debra Meloy Elmegreen\affil{Vassar College, Dept. of Physics \& Astronomy, 
Box 745, Poughkeepsie, NY 12604; elmegreen@vassar.edu}}

\author{Bruce G. Elmegreen \affil{IBM Research Division, T.J. Watson
Research Center, P.O. Box 218, Yorktown Heights, NY 10598, USA,
bge@watson.ibm.com} } 

\author{Kate S. Eberwein \affil{Vassar College, Dept. of Physics \& Astronomy, 
Box 745, Poughkeepsie, NY 12604, kaeberwein@yahoo.com } } 

\begin{abstract} 
The structure of dust spirals in the nuclei of the SAab-type Liner
galaxies NGC 4450 and NGC 4736 is studied using archival HST PC images.
The spirals are typically only several hundredths of a magnitude fainter
than the neighboring disks, so unsharp mask techniques are used to
highlight them.  The ambient extinction is estimated to be less than
0.1 mag from the intensity decrements of the dust features and from the
spiral surface filling factor, which is about constant for all radii and
sizes.  The nuclear dust spirals differ from main-disk spirals in several
respects: the nuclear spirals have no associated star formation, they are
very irregular with both trailing and leading components that often cross,
they become darker as they approach the center, they completely fill the
inner disks with a constant areal density, making the number of distinct
spirals (the azimuthal wavenumber $m$) increase linearly with radius,
and their number decreases with increasing arm width as a power law.
Fourier transform power spectra of the spirals, taken in the azimuthal
direction, show a power law behavior with a slope of $-5/3$ over the range
of frequencies where the power stands above the pixel noise.  This is
the same slope as that found for the one-dimensional power spectra of
HI emission in the Large Magellanic Cloud, and also the slope expected
for a thin turbulent disk.  All of these properties suggest that the dust
spirals are a manifestation of acoustic turbulence in the inner gas disks
of these galaxies.  Such turbulence should dissipate orbital energy and
transfer angular momentum outward, leading to a steady accretion of gas
toward the nucleus.  \end{abstract}


\section{Introduction} 

Dust spirals in the inner kpc regions of galaxies reveal a source of
compression that can affect the angular momentum distribution of the gas
and possibly drive accretion to an AGN. Barred galaxies with an inner
Lindblad resonance tend to have two long and symmetric dust spirals near
the resonance that are a continuation of the leading-edge dust lanes in
the bar (Athanassoula 1992). Many barred
galaxies have ILR rings too (Buta \& Crocker 1993; P\'erez-Ram\'irez et
al. 2000; Knapen et al. 2000).  Non-barred galaxies, galaxies without
an ILR (e.g., late Hubble types), and regions of barred galaxies inside
their ILRs can have more irregular dust spirals.

Whether the nuclear spirals are regular or irregular, their presence
in optical images suggests a density variation that
is at least a factor of 2 and therefore likely to involve shocks. These
shocks are oblique for azimuthal flows, so the gas will experience a
torque when it enters a spiral and an opposite torque when it
leaves. In cases where the gas moves faster than the spirals, the net
torque is negative and the gas loses angular momentum and energy at the
shock, causing it to spiral inward. If the spirals move faster than the
gas, as might be the case outside a fast nuclear bar, then the gas gains
angular momentum and may gradually move out.

We are interested in the source of compression for the irregular nuclear
dust spirals that appear in non-barred galaxies or inside the ILRs of
some barred galaxies. We have proposed that some of these are caused by
random sonic noise that amplifies weakly as it propagates toward the
center (Elmegreen et al. 1998; Montenegro, Yuan, \& Elmegreen 1999;
Englmaier \& Shlosman 2000). A
signature of this process is an irregularity of structure, a wide range
of pitch angles reflecting different times of origin and different wave
propagation directions, a tendency to fill the volume available with a
spiral separation comparable to the epicyclic radius for motions at the
sound speed, and a general trend of increasing density with decreasing
radius for each spiral. Because the velocities inside these spirals
cannot yet be measured, there is no way to be sure they are driving
inflow. Nevertheless, if any of the spirals come close to
the nucleus, then such accretion would seem to be likely.

Nuclear dust spirals and clouds have been studied previously using Hubble
Space Telescope (HST) data. Van
Dokkum \& Franx (1995) found them on WFPC V-band images of early
type galaxies.  Malkan, Gorjian, \& Tam (1998) did a large
snapshot survey of active galaxies and classified the nuclear morphology
including the dust. 
Elmegreen et al. (1998) noted dust spirals in WFPC2 images
of the interacting galaxy NGC 2207 and proposed they might drive
accretion to the nucleus. 
Regan \& Mulchaey (1999) found nuclear dust spirals in 6 AGN galaxies
using WFPC2 and NICMOS images, and Martini \& Pogge (1999) found them
with HST data in 20 Seyferts; both studies concluded that the spirals
could drive accretion. 
Ferruit, Wilson \& Mulchaey (2000) studied 12 early
type Seyfert galaxies in more detail and noted the dust spirals too.
Tomita et al. (2000) used HST archival images to
study dust features in E and S0 galaxies, but did not comment on spirals
specifically. 
Tran et al. (2001) studied nuclear dust in elliptical galaxies.

Here we measure the extinction and structural properties of irregular dust
spirals in the central kpc of 
two LINER galaxies, NGC 4450 and NGC 4736. Their Hubble types
are about the same, SA(s)ab and RSA(r)ab (de Vaucouleurs et al. 1991),
and their distances are taken to be 16.8 and 4.3 Mpc (Tully 1988).

\section{Observations}

Hubble Space Telescope preview images of the galaxies NGC 4450 and NGC
4736 are shown in Figure 1. These images were obtained from the HST
archive web site and are degraded in resolution by a factor of 2 from
the full HST data; they are from proposals 5375 by V. Rubin and 5741
by J. Westphal, respectively. For our analysis of dust structure, we
used the original Planetary Camera (PC) images with 
$0.0455^{\prime\prime}$ per pixel, corresponding to 3.7 pc and
0.95 pc per pixel, respectively.  The PC fields of view cover 3000 pc and
760 pc.  Multiple images in the V passband (F555W filter) were combined
to remove cosmic ray tracks and bad pixels.
Figure 1 shows the full WFPC2 image to illustrate the outer
disk morphologies, their position angles,
and the orientation of the HST images relative to north.

\subsection{Dust Absorption}

Intensity decrements for dust cloud absorption were measured for many
features using radial and azimuthal profiles on the PC images. Figure
2 shows a sample profile from the V-band image of NGC 4736
taken along the strip shown in Figure 3 
(which uses the G4-G12 enhanced image from Fig. 6 below). Most of the
intensity dips deeper than a few hundredths of a magnitude 
are dust features, as
can be determined by comparing Figures 2 and 3. The
thinner features may be verified as dust also by comparing Figure 2
with the G1-G3 and G2-G6 images in Figure 6.

Intensity decrements do
not measure the cloud extinction because there is likely to be a
smooth intensity contribution from foreground starlight. If dust
with opacity $\tau$ in the midplane is next to an intercloud medium
with opacity $\tau_0$, then the cloud to intercloud intensity ratio is
$\left(1+e^{-\tau}\right)/\left(1+e^{-\tau_0}\right)$. For $\tau >\tau_0$,
this ratio is between 0.5 and 1, making the maximum cloud contrast 0.75
mag. Thus the magnitude difference is not a good measure of intrinsic
extinction unless $\tau$ is small and $\tau_0$ is either independently
known or close to zero. An intensity ratio lower than 0.5 corresponds to
an opaque dust cloud on the near side of the galaxy.

Radiative transfer models of the intensity distributions in several
passbands are necessary to determine the intrinsic opacities of these
clouds, but the present HST data were not acquired for this purpose and
the wavelength coverage and integration times are not adequate for
a proper analysis.  Dust spirals in another, similar, nuclear region
were analyzed by radiative transfer 
previously using HST data in V and I passbands (Elmegreen et
al. 1998). We found that typical extinctions are several magnitudes
inside the darkest clouds, and that the average extinction in the
inner disk is so low that the gas is not likely to be self-gravitating.
We expect the same situation here because of a lack of star formation
associated with the dust.  Radiative transfer models of dust features
for main galaxy disks are more common than for nuclear disks (e.g.,
Howk \& Savage 1997; Kuchinski et al. 1998; Regan 2000).  Galaxy
extinctions have also been measured using background sources
(e.g., Pizagno \& Rix 1998;
White, Keel, \& Conselice 2000;  Keel \& White 2001;
Elmegreen et al. 2001)

The typical absorption trough for a dust feature in Figure 2 is several
hundredths of a magnitude in V band. The deep feature indicated is part
of a long dust spiral that covers a factor of two in radius and then
branches into two thinner spirals in the inner part (see Fig. 3). Figure 4 shows
the magnitude difference for this spiral and several others in the two
galaxies as a function of radius.  These differences were measured by
comparing the intensities of the dust lanes to the adjacent interdust
regions at several radii. Estimated measuring uncertainties are 0.002 mag.
Each line in Figure 4 is a different spiral; the top dashed line
corresponds to the dust feature shown in Figure 2. The darkness of the dust
generally increases inward, suggesting that either the density of the
spirals or their thickness relative to the stellar disk increases inward.

\subsection{Image Enhancement, Fractal Structure, and Dust Opacity}

The spiral dust structures are very faint. They require enhancement
with unsharp masks or other filtering techniques. Figures 5 and 6 show
composites of four filtered images for each galaxy. Each filter isolates
structures within a range of scales spanning a factor of 3. The top left
panels contain unsharp-masked images made by subtracting a smoothed
version from the original image, using a smoothing function that is a
Gaussian with a dispersion of $\sigma=3$ pixels. This image shows the
smallest structures, from 1 to 3 pixels. In the top right, a smoothed
image with a $\sigma=6$ px Gaussian was subtracted from a smoothed image
with a $\sigma=2$ px Gaussian (this will be called a G2-G6 image). This
shows features with scales between 2 px and 6 px. The bottom left and
right images subtract $\sigma=12$ px from $\sigma=4$ px convolutions
(G4-G12) and $\sigma=24$ px from $\sigma=8$ px convolutions (G8-G24),
respectively. Each image has considerable structure, indicating that
clouds span a wide range of scales.

The fractal dimension of multiscale structure can be determined from box
counting or cloud counting techniques, plotting the number of structures
as a function of scale in log-log coordinates (Mandelbrot 1982; Westpfahl
et al. 1999).  Because we have already isolated certain scales with
each filter, the cloud counting results on the subtracted images will
give the same slope as the derivative with respect to length of the
box counting results on the original images.  Recall that box counting
includes all structures larger than the scale of the box. The fractal
dimension $D$ is the slope of the function $N(>S)$ versus $S$ on log-log
coordinates for number of structures larger than $S$ determined by the box
counting technique. If we write $N(>S)=\int_S^\infty n(S)dS$ for number of
structures $n(S)$ between $S$ and $S+dS$, then $n(S)dS\propto S^{-D-1}dS$
and $n(S)d\log S \propto S^{-D}d\log S$ for linear and log intervals
of size, respectively. Thus the slope on a log-log plot of the number
of clouds in each interval of filter size, for logarithmically spaced
intervals like we have in Figures 5 and 6, gives the fractal dimension
of the cloud structures. We choose the filter technique rather than the
box technique so we can see the nature of the structure on each scale,
whether it is filamentary or shell-like, for example. Figures 5 and 6
indicate that it is filamentary on all the resolved scales.

Figure 7 shows the number of spirals intercepted by deprojected
azimuthal scans on the G2-G6 image plotted as a function of radius.
The radii used for the measurements are 22, 45, 67, 90, and 112 pixels.
A deprojected azimuthal scan of this image showing average counts for the
radial interval from 66 to 68 pixels is shown in Figure 8.  The number of
clouds plotted in Figure 7 at this radius, 22 clouds, is shown again in
Figure 8, with each cloud indicated by an arrow. The full circle of the
azimuth in Figure 8 corresponds to 421 pixels.  Generally the cloud count
can be done easily by eye on a computer screen down to the brightness
limit of the image.  The number of dark features in an azimuthal scan
is always about one-third of the radius, which makes the azimuthal
spacing between them $\sim6\pi$ pixels.  Because the average cloud size
on this image is $\sim4$ px, the clouds fill $4/\left(6\pi\right)=21$\%
of the circumference.

This spiral filling factor makes sense given the extinction contrast.
If the spirals formed by the compression of a fraction $f$ of the gas
and the spiral spacing is $C\sim5$ times the spiral width ($C$ is the
inverse of the filling factor given above), then the column density of
each spiral is $(C-1)f+1$ times the ambient average.  This column density
has a contribution from uncompressed material that was at the spiral
position already, which is the fraction $1-f$ of the ambient column
density, plus the compressed material from the whole volume out to the
next spiral, which is the amount $fC$ times the ambient column density.
The sum is $1-f+fC=(C-1)f+1$ times the ambient column density, as above.
The assumption here is that not all of the interstellar medium will be
compressed by the turbulence.  A warm or hot component, with a thermal
speed larger than the mean turbulent speed, should flow around the moving
clouds and not compress easily.  Thus $f$ may be considered the mass
fraction of the gas that is compressible by the prevailing turbulence.

In this situation,
suppose the spiral opacity is $\tau$, the
interspiral opacity is $\tau_0$, and the ambient average before the
compression was $\tau_a$, all from gas concentrated in the midplane.
Then the compression gives $\tau=\left[\left(C-1\right)f+1\right]\tau_a$,
the conservation of mass gives $\tau+\left(C-1\right)\tau_0=C\tau_a$,
and the density contrast in magnitudes, $\Delta m$, gives \begin{equation}
{{1+e^{-\tau}}\over{1+e^{-\tau_0}}}=10^{-0.4\Delta m}
\end{equation}
These equations can be solved for the three opacities given $C$,
$f$, and $\Delta m$. This solution has a simple expression in the
limit of small $\Delta m$.  In that case, there are two solutions
for $\tau$, $\tau_0$ and $\tau_a$, one where the opacities
are very large so hardly any background light gets through,
and another where the opacities are very small.  The relatively
smooth continuation of the outer disk starlight into the inner
region implies the latter.  Then in the limit of small $\Delta m$,
\begin{equation} \tau_a\sim 0.8\ln10\Delta m/\left(fC\right).
\end{equation} 
For $C\sim5$ and $f\sim1/2$, this gives $\tau_0\sim0.7\Delta m$ on
average.  Considering $\Delta m\sim0.04$ from Figure 4, the ambient
opacity of $\tau_a\sim0.03$ corresponds to a mass column density of
$\sim0.5$ M$_\odot$ pc$^{-2}$.  This result is $\sim10$ times smaller
than the average extinction in the center of NGC 2207, which was about
0.6 mag (Elmegreen et al. 1998), but the dust features here are ten times
fainter than in NGC 2207, where a prominent dust lane used for the opacity
measurement had a V-band intensity contrast of 0.3 mag.  Evidently, the
column densities in NGC 4450 and NGC 4736 are too small for the inner
gas disks to be significantly self-gravitating; for reasonable epicyclic
frequency and velocity dispersion, the Toomre stability parameter $Q$
exceeds 100.

We can do this same exercise for spirals of different sizes.  The top
part of Figure 9 shows the number of dust spirals counted in different
azimuthal scans versus the size of the Gaussian filter, plotting the
G1-G3 filter combination as a point with an average Gaussian radius
of 2, for example. The radii used for the azimuthal scans are 45, 67,
90 and 112 pixels.  The number of dust spirals decreases approximately
inversely with Gaussian filter radius for all galactic radii, suggesting
again that the spirals are packed in tightly, even on a variety of scales.
If the spacing between spirals in the azimuthal direction is $\lambda$,
and this spacing is the factor $C$ times the spiral width, $W$, then the
number $N$ of spirals at radius $R$ is $2\pi R/\lambda$ and the inverse
of the filling factor is $C=2\pi R/\left(NW\right)$.  This inverse
filling factor, $C$, is shown in the bottom of Figure 9.  For all but
the smallest features, which are probably undercounted, $C\sim3-5$
for all sizes and radii.  This is the same value of $C$ that was used
in the previous discussion of $\tau_a$.  Thus the spirals are likely to
be made on a wide variety of scales by {\it compressions} of a factor
of $C\sim3-5$ from an ambient medium with an opacity slightly less than
0.1 mag. We suggest below that the mean gas 
density could be $\sim1.5$ cm$^{-3}$.

The top part of Figure 9 may also be used to determine a fractal
dimension, which equals the slope according to the discussion at the
beginning of this subsection.  The number of clouds decreases a little
less rapidly than the inverse of the smoothing scale $\sigma$, giving
an apparent fractal dimension slightly less than 1.  If we write the
dependence of the number of counted clouds on the cloud size as a power
law, $\sigma^{-D}$, for logarithmic intervals of $\sigma$, then $D$ is the
fractal dimension. From Figure 9, $D$ equals $0.79\pm0.16$ for NGC 4450
and $0.51\pm0.04$ for NGC 4736, considering Gaussian radii from 4 to 32.

\subsection{Fourier Transform Power Spectra}

Fourier transform power spectra of the azimuthal intensity profiles
at ten radii in each galaxy are shown in Figure 10.  The radii $R$ were
chosen so that $2\pi R$ equals an allowable number of pixels in a Fast
Fourier Transform done with the SRCFT subroutine in the IBM ESSL software
package. These numbers give radii $R=$ 33, 57, 81, 107, 134, 153, 183,
204, 229, and 244 pixels.

The power spectra are the sums of the squared real and imaginary parts
of the FFTs.  These are one-dimensional FFTs, taken from the azimuthal
intensity traces along the deprojected images.  Deprojection was made
with a position angle of $7^\circ$ and an inclination of $42.2^\circ$
for NGC 4450 on the HST image, and with a position angle and inclination
of $-2^\circ$ and $35.6^\circ$ for NGC 4736.  The inclinations were taken
from the Third Reference Catalogue of Bright Galaxies (de Vaucouleurs
et al., 1991).  The position angles were measured from the HST images
of the nuclear regions and differ from the position angles of the outer
disks.  In terms of the number of degrees counter-clockwise from North,
the position angles of the nuclear-disk major axes are $\sim1^\circ$
and $\sim21^\circ$ for NGC 4450 and NGC 4736, respectively.

The projected ellipses used for the intensity traces are shown on the
G4-G12 images in Figure 11.  Each intensity trace is an average over 9
deprojected azimuthal scans spaced by one pixel and centered on one of
the radii $R$ given above.  Each pixel used for an azimuthal average is
blackened in Figure 11.

The power spectra are illustrated in Figure 10 by stacking the different
radii on top of each other with the smallest radius at the bottom.
The vertical scale corresponds to a factor of 10 in the power spectrum
for each tic mark; the absolute scale is arbitrary.  The solid,
dashed, and dot-dashed lines have slopes of $-5/3$, $-1$, and $-8/3$.
Solid lines with a slope of $-5/3$ are drawn through the low frequency
portions of each power spectrum to illustrate the general trend.

The power spectra all have slopes of about $-5/3$ at low frequency. The
slopes flatten at high frequency to a slope of about $-1$ or flatter.
The low frequency power spectrum could be steeper than $-5/3$ at the
smallest radius, possibly as steep as $-8/3$, as shown by the dot-dashed line.

Turbulence gives a one-dimensional power spectrum with a slope of $-5/3$
for structures that are thinner on the line-of-sight than the transverse
wavelength, and a slope of $-8/3$ for structures that are thicker on
the line-of-sight than the wavelength (Lazarian \& Pogosyan 2000).
Two-dimensional power spectra have slopes for these two cases that
are steeper by one, i.e., $-8/3$ and $-11/3$.  The azimuthal scan
results here, giving slopes of $-5/3$ at small spatial frequencies,
are consistent with turbulent structures that are wider than the galaxy
disk scale height.

The same slope of $-5/3$ was found for power spectra of azimuthal scans
of the HI intensity of the Large Magellanic Cloud (Elmegreen, Kim, \&
Staveley-Smith 2001).  Thus the origin of the structure is probably the
same.  Power law power spectra have also been found in other interstellar
emission line surveys, using HI (Crovisier \& Dickey 1983; Green 1993)
and CO (Stutzki et al. 1998) in the Milky Way and HI in the Small Magellanic
Cloud (Stanimirovic et al. 1999).  The slopes are always consistent with
turbulence as the origin of the structure.

The LMC showed a turnover in the power spectrum at high spatial frequency
in both the azimuthal and linear scans, going from slopes of $-5/3$
to $-8/3$ at high spatial frequency, and it showed a similar turnover
in the two-dimensional power spectrum of the whole disk, going from
$-8/3$ to $-11/3$ at high frequency.  These turnovers were proposed
to correspond to the transition from structures that are larger in the
transverse direction than the galaxy thickness to structures that are
smaller than the galaxy thickness (Elmegreen, Kim, \& Staveley-Smith
2001), as predicted in the general case by Lazarian \& Pogosyan (2000).
Elmegreen et al. also found that LMC thickness increased with radius.
This result was confirmed by Padoan et al. (2001), who found a similar
thickness-dependent feature in the spectral correlation function of HI
emission from the LMC, and used this feature to map the inferred disk
thickness around the plane.  Analogous turnovers in the power spectrum
are not seen for the nuclear dust spirals studied here, presumably because the
noise level is too high.  This problem is suggested by the flattening
of the power spectra at intermediate to high frequencies -- before the turnover
can occur.  Longer integration times on these sources could
possibly get the noise down to a level where the power spectrum
turns over.  Then the line-of-sight disk thicknesses could be measured.
All we can say here is that the thickness is less than $\sim50$ pc in
NGC 4450 and less than $\sim10$ pc in NGC 4736. These limits are not
unreasonable because they can still give the inner gas disks about the
same aspect ratios as a typical main galaxy gas disk.  When combined
with the average extinction estimate given in the previous subsection,
$A_V\sim0.03$ mag, a disk thickness of $\sim10$ pc corresponds to an
average gas density of $\sim 1.5$ cm$^{-3}$, which is about 
the same as that in the Solar neighborhood.

The morphologies of the dust structures in the nuclear regions of NGC 4450
and NGC 4736 differ from the morphologies of the gas emission structures
in the Milky Way and Magellanic Clouds surveys.  In the nuclear regions,
the dust is mostly in the form of spirals, which show a power-law power
spectrum in the azimuthal direction and a gradient of intensity in the
radial direction (cf. Fig. 4). For the HI and CO emission line surveys,
the structures are more blob-like or shell-like, with essentially no
spirals, even in the whole LMC and SMC. Evidently, the origin of these
gaseous structures is the same, i.e., random and persistent compressions
from turbulence, but the nuclear regions of NGC 4450 and NGC 4736 have
a lot of shear also, and this shear distorts the compressed regions
into spirals.  The LMC and SMC have relatively little shear, as do local
interstellar clouds.

Numerical simulations of this shearing effect for gravitating media were
in Toomre \& Kaljnas (1991), Huber \& Pfenniger (1999), Wada \& Norman
(1999), and Semelin \& Combes (2000).  Pressures from star formation were
also considered by Wada, Spaans, \& Kim (2000) and Wada \& Norman (2001).
The present case differs from these others slightly because the nuclear
dust structures here are mostly without star formation and they are
probably also without significant self-gravity.  The origin of the
turbulence is unknown, although a mild, pressure-driven instability was
suggested by Montenegro et al. (1999).

\section{Discussion}

The nuclear spirals in NGC 4450 and NGC 4736 vaguely resemble the
outer spirals, which are somewhat flocculent in each case, but the
nuclear spirals do
not continue smoothly from the outer spirals and there are important
structural
differences. The B-band image of NGC 4450 in the Carnegie Atlas of
Galaxies (Sandage \& Bedke 1994) shows two long dust spirals in the main
disk, along with some flocculent structure; the stellar spiral arms are
smooth. In contrast, the nuclear region has no stellar arms and at least
7 prominent dust arms, some with pitch angles as high as $45^\circ$ and some
crossing each other.  The eastern side of the nuclear region shows more
dust than the western side because of the galaxy's inclination. Some small
dust feathers extend nearly radially from the center toward the south,
reminiscent of jets. HST spectral observations by Ho et al. (2000) reveal
double-peaked line profiles with high velocity wings, characteristic of
accretion disk activity observed in other LINERs.

NGC 4736 is an early-type galaxy with an outer ring and a circumnuclear
starburst ring.  Its main disk structure is flocculent and defined
primarily by the dust.  Sandage \& Bedke's B-band print shows the inner
disk structure as composed of many tightly-wrapped arms, but the central
region is saturated in the reproduction. Waller et al. (2001) present
UIT UV and ground-based R-band images of the central regions, including
an unsharp-masked image showing the complicated flocculent structure of
the main disk.  They also show an HST FOC image of the main nuclear dust
arms. Ground-based NIR observations by Mollenhoff, Matthias, \& Gerhard
(1995) suggested a weak stellar bar with a length of $20^{\prime\prime}$,
which was also noted by Maoz et al. (1995) from an HST FOC image.
The bar was observed in CO by Sakamoto et al. (1999) and Wong \& Blitz
(2000).  In the HST image, the region corresponding to the bar shows up
as an elongated disk with a position angle nearly perpendicular to the
major axis of the galaxy.  The structure inside the circumnuclear ring,
which is really two tightly wrapped arms, consists of a dozen dust arms
within a radius of 50 pc from the center, branching to dozens more dust
spirals out to 200 pc. The nuclear dust spirals are not attached to the
main inner disk dust spirals.

The nuclear dust in NGC 4450 and NGC 4736 has several characteristics
that differ from spiral arms and dust clouds in main galaxy disks.  These are:

\begin{itemize}

\item The nuclear dust spirals shown here have no associated star
formation.  Other nuclear spirals in different galaxies have star
formation (e.g., Coma D15 in Caldwell, Rose, \& Dendy 1999), so the
gaseous nature here is not universal. The lack of star formation suggests
that the inner gas disks in NGC 4450 and NGC 4736 are not strongly
self-gravitating. The same was true for NGC 2207 (Elmegreen et al. 1998)
and for several other inner disks in the study by Martini \& Pogge (1999).
Our opacity estimate in Section 2.2 is also consistent with this. 

\item The nuclear dust is in the form of spiral arms of various pitch
angles, widths, and lengths. Some of the arms are trailing, a few are
leading, and many cross each other.  This pattern is generally more
irregular than main disk flocculent arms (see atlas in Elmegreen 1981).
Main disk flocculent arms are rarely leading. They generally do not cross
each other; if they branch into spurs, then this branching is toward larger
radii (Elmegreen 1980). They also have star formation that gives
them a thicker, more patchy quality, rather than a filamentary quality.

\item The nuclear dust spirals in NGC 4450 and NGC 4736 have decreasing contrast with
increasing radius. Ambient dust extinctions generally decrease with
galactocentric distance because of the exponential distribution of gas
column density. The flocculent galaxy NGC 5055 has such a decrease, for
example, as measured by the extinctions of OB associations (Acarreta et
al. 1996). Nuclear dust spirals are not just ambient extinctions,
however. They are morphologically more similar to main disk spiral arms
than diffuse cloud extinctions because they are organized and most
likely formed by compressive processes in the presence of shear. From
this point of view, nuclear dust spirals should be compared to main disk
spirals, and then the radial decrease in nuclear spiral amplitude is
unusual. Density wave spirals in non-barred galaxies tend to get
stronger with increasing radius, out to at least the corotation zone
(Elmegreen \& Elmegreen 1984; Elmegreen et al. 1996). 
The unusual result that nuclear dust
spirals get weaker with radius is presumably the result of crowding near
the center for waves that move inward, as predicted for solutions to
Bessel's wave equation (Elmegreen et al. 1998; Montenegro et al. 1999).
This is a different dynamical situation than for main disk spiral
arms, for which the curvature terms ($\propto 1/kr$ for wavenumber $k$)
in the wave equations can usually be ignored (e.g., Bertin et al. 1989).

\item The number density of dust features is about constant with radius,
indicating that the inner disk is completely filled with structure. This
is unlike the situation for main galaxy disks which often have a small
number of arms (e.g., 2-5) that get further apart with
radius. The nuclear spirals also have some indication of a hierarchical
or fractal structure because of a non-integer slope of the size
distribution, examined with unsharp masks. This size distribution is
approximately a power law with a slope in the range from 0.5 to 0.8.
This power law is reminiscent of other properties of interstellar clouds
formed by turbulence, but complicated in this case by the effects of
shear, which make spirals rather than clumps, and by the Coriolis force,
which resists turbulent motions on large scales. The dust features are
also very weak, and the smaller clouds, as well as those further from the
center, are difficult to see above the pixel noise. Fractal structure in
the dust of another galaxy was also found by Keel \& White (2001)
using a background illumination source. Fractal structure in galactic
clouds is well known (e.g., Falgarone, Phillips, \& Walker 1991).

\item Fourier transform power spectra in the azimuthal direction show
the characteristic signature of turbulence compression, which is a power
law slope of $-5/3$ for one-dimensional structures that are larger than
the line-of-sight thickness of the galaxy disk. This makes the nuclear
dust features studied here resemble the HI clouds in the LMC, with the
important difference that the nuclear clouds are spiral filaments,
presumably affected by shear, and the LMC clouds are globular and
shell-like in a low shear environment.

\end{itemize}

There is no direct evidence in our observations for accretion driven
by the dust spirals. Radial velocities will have to be measured to
determine this.  However, the increase in dust opacity with decreasing
radius for the main spirals is consistent with the amplification
that is expected for inward motions.  In that case, the spirals
could drive nuclear accretion. 

\acknowledgments 
D.M.E. and B.G.E. gratefully acknowledge HST archival grant HST-AR-09197.
We also acknowledge useful comments by the referee.

\newpage

\begin{figure}
\caption{NGC 4450 (left) and NGC 4736 (right)
V-band images from HST WFPC2. 
(See jpeg files in astro-ph.) } 
\end{figure}

\clearpage

\begin{figure}
\plotone{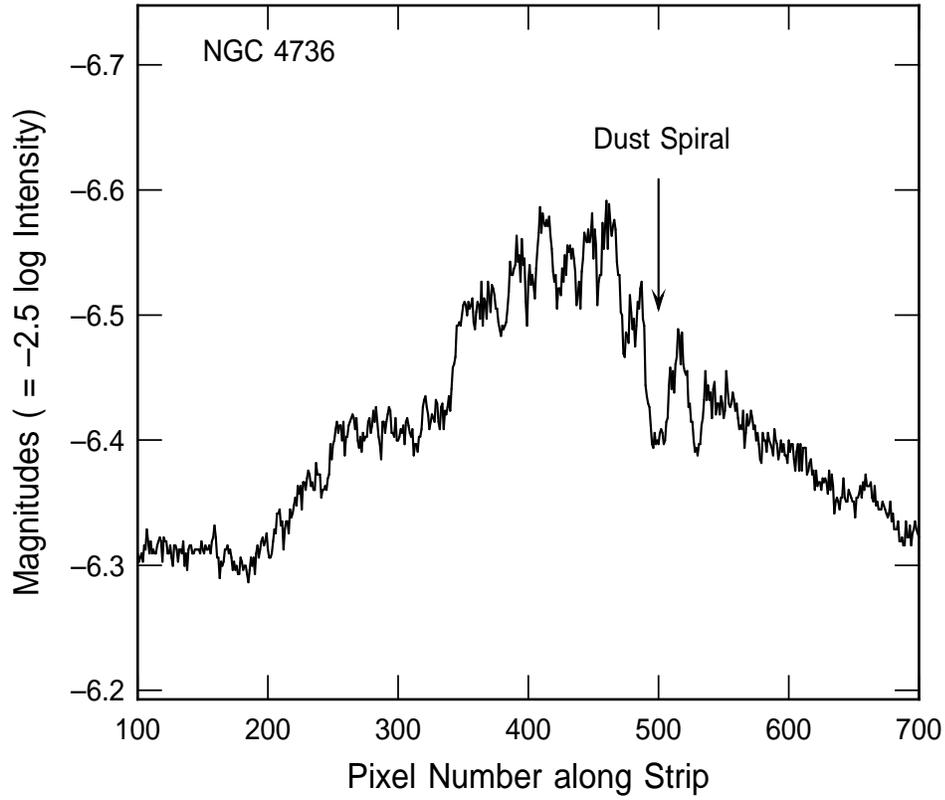}
\caption{An intensity trace across a dust spiral in NGC 4736 showing
an absorption dip, which is labeled, 
and other dust features.
The maximum of intensity in the center of the scan is from the exponential disk.} 
\end{figure}
\clearpage

\begin{figure}
\caption{The 600-pixel long
scan used for Fig. 2 is shown superposed on the G4-G12 image from
Fig. 6.  The dust is dark in this image.
(See jpeg file in astro-ph.) }
\end{figure} 

\begin{figure}
\plotone{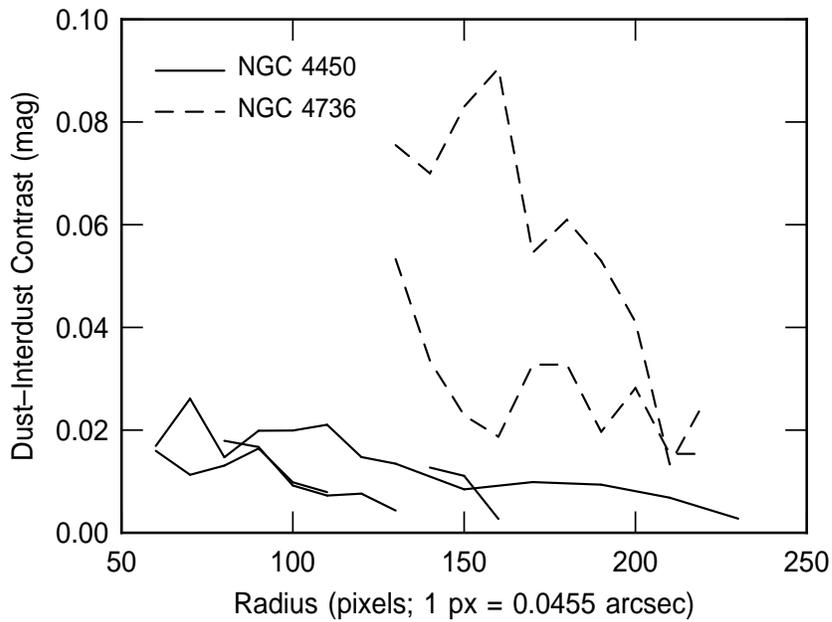}  
\caption{The magnitude differences between the dust features and
the neighboring interdust regions are plotted versus the
distance from the galaxy centers for four dust spirals in NGC 4450
and two dust spirals in NGC 4736. The dust spirals tend to get darker toward
the center. } 
\end{figure}

\clearpage

\begin{figure}
\caption{Unsharp mask images of NGC 4450, made by subtracting
two Gaussian smoothed images, Gn-Gm, for Gaussian dispersions n and m, 
measured in pixels. The top left
panel shows G1-G3; top right: G2-G6; bottom left: G4-G12, and bottom right:
G8-G24. The number of pixels along the horizontal edge of each
image is 725, corresponding to 2700 pc. 
(See jpeg file in astro-ph.) } 
\end{figure} 

\clearpage

\begin{figure}
\caption{Unsharp mask images of NGC 4736, as in Fig. 5.
The number of pixels along the horizontal edge of each
image is 590, corresponding to 560 pc. (See jpeg file in astro-ph.)} 
\end{figure} 

\clearpage

\begin{figure}
\plotone{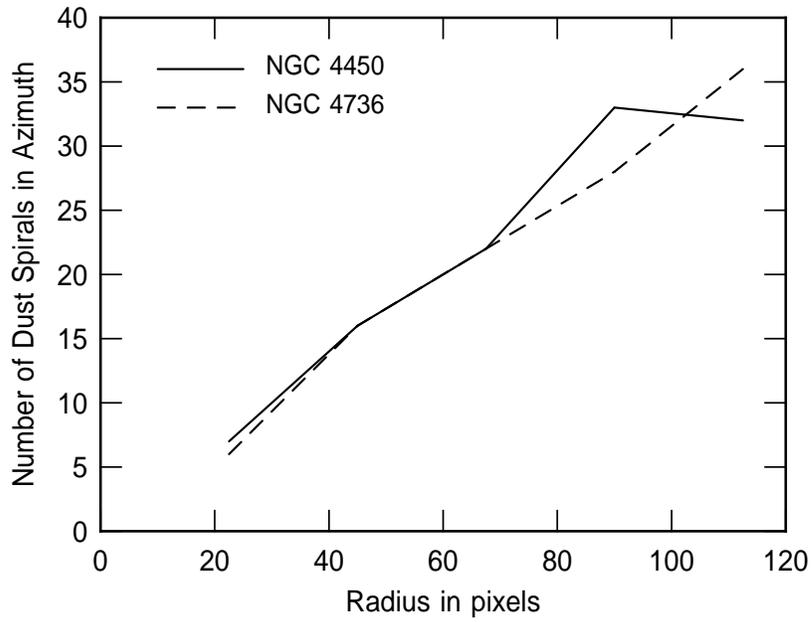}  
\caption{The numbers of distinct dust spirals in azimuthal scans
on the G2-G6 images are 
plotted versus the radii. The numbers equal approximately
one-third the radii, indicating that the filling factor for the
dust is about 0.2 on the scale of $\sim4$ pixels.} 
\end{figure} 

\clearpage

\begin{figure}
\plotone{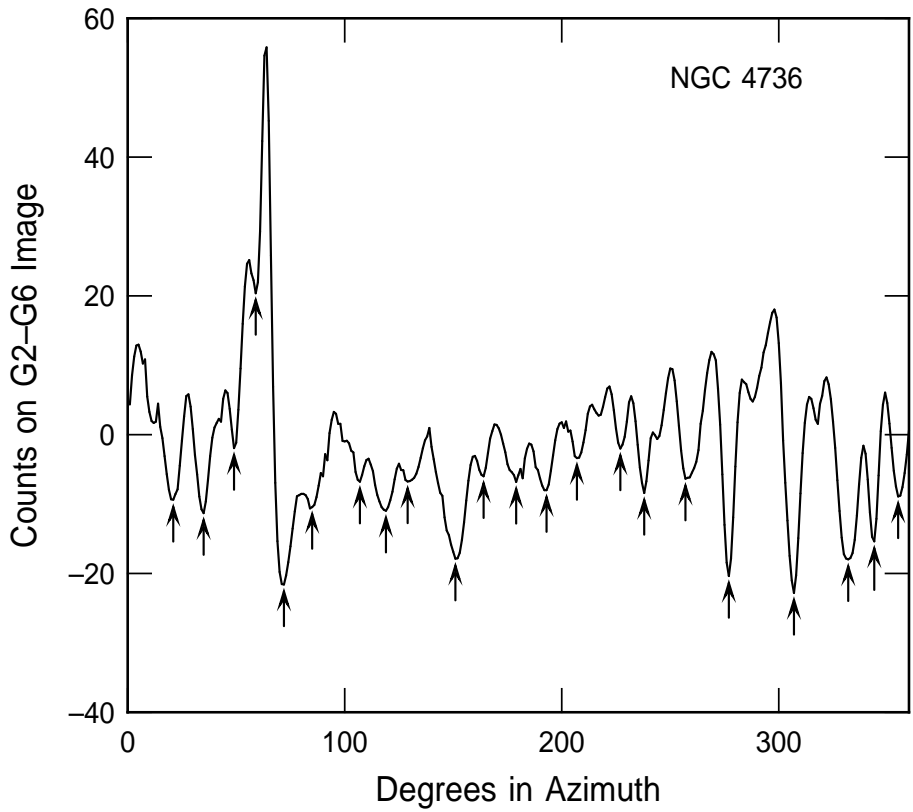}  
\caption{A deprojected azimuthal scan at a
radius of 67 pixels for the G2-G6 image of 
NGC 4836 showing dust features as negative counts.  The
22 features counted for Figure 7 are indicated. 
The full azimuth of $360^\circ$ plotted here corresponds to 421 pixels, so 
the average spacing between these clouds is 421/22=19 pixels,
which is close to the value of $6\pi$ discussed in the text. 
}
\end{figure} 

\clearpage

\begin{figure}
\plotone{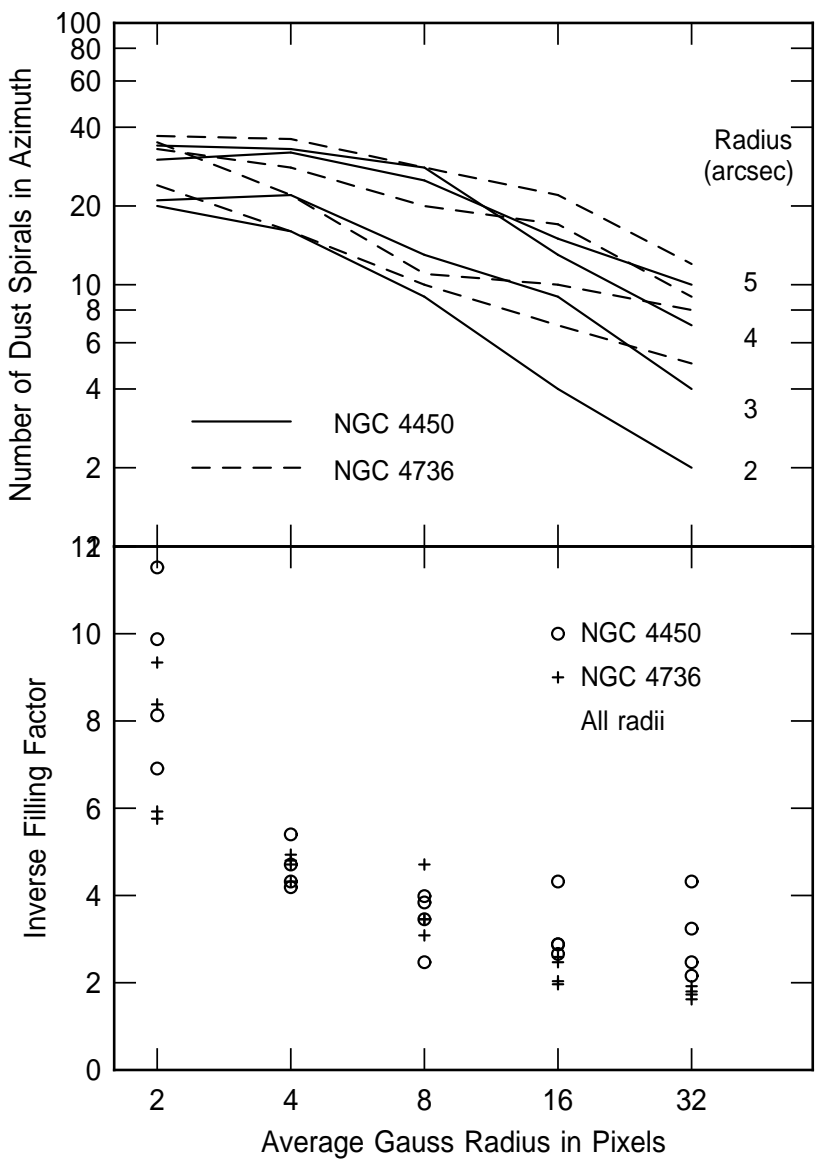}  
\caption{(top) The number of distinct dust spirals in azimuthal scans
of different radii is shown as a function of the Gaussian dispersion
of the unsharp mask, which is the characteristic feature size. 
The radii for the scans are 45, 67, 90, and 112 pixels. 
The number of spirals decreases approximately as a power law over
a factor of ten in size. The slope of this power law is
the fractal dimension. 
(bottom) The ratio of the azimuthal spacing between 
the spirals to the spiral width given by the average Gaussian smoothing radius.
This ratio is the inverse of the areal filling factor of the spirals, 
and it is about constant, equal to $\sim4$ for all radii and
spiral arm widths larger than $\sim4$ pixels.   }
\end{figure} 

\begin{figure}
\plotone{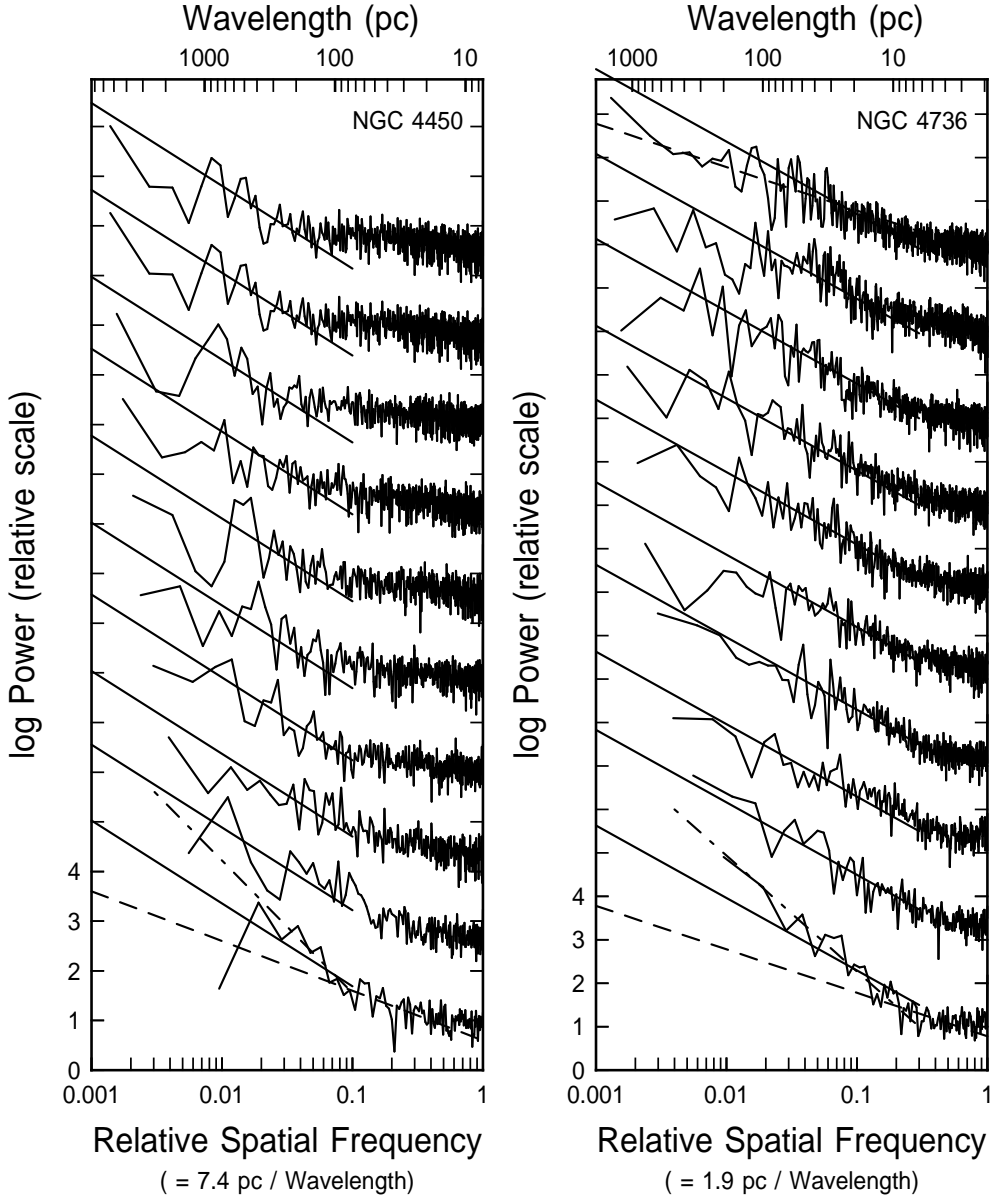}  
\caption{Fourier Transform power spectra of total intensity along
the azimuthal scans shown in 
Fig. 10.  The scans are stacked in order of 
increasing radius in the galaxy.  The solid lines drawn through each scan
have a slope of $-5/3$, the dashed line
at the bottom and top has a slope of $-1$, and the dot-dashed line at the
bottom has a slope of $-8/3$.   The similarity between the one-dimensional power spectra of 
dust spirals in these galaxies, the 1D power spectrum of HI emission on large scales
in the
Large Magellanic Cloud, and the 1D power spectrum of two-dimensional turbulence
suggests that the spirals result from turbulent compression in a relatively thin disk.}
\end{figure} 

\begin{figure}
\caption{The projected azimuthal scans used for the power spectra are shown superposed on the
G4-G12 images.
(See jpeg files in astro-ph.) } 
\end{figure}

\end{document}